\documentclass{iaus}
\usepackage{graphicx}
\usepackage{enumerate}

\newcommand{\vc}[1]{\mbox{\boldmath{$#1$}}}
\newcommand{\de}{\mathrm{d}}
\newcommand{\dpa}{\partial}

\newcommand{\nab}{\vc{\nabla}}

\newcommand{\ii}{\mathrm i}
\DeclareMathSymbol{\varOmega}{\mathord}{letters}{"0A}
\DeclareMathSymbol{\varSigma}{\mathord}{letters}{"06}
\DeclareMathSymbol{\varPsi}{\mathord}{letters}{"09}
\DeclareMathSymbol{\varPhi}{\mathord}{letters}{"08}
\DeclareMathSymbol{\varGamma}{\mathord}{letters}{"00}

\newcommand{\apj}{Astrophys.~J.}
\newcommand{\apjl}{Astrophys.~J.}
\newcommand{\mnras}{Mon.~Not.~R.~Astron.~Soc.}
\newcommand{\aap}{Astron.~Astrophys.}

\newcommand{\nat}{Nature}

\newcommand{\araa}{ARA\&A}

\title[The role of magnetic fields for planetary formation]
{The role of magnetic fields\\for planetary formation}

\author[A. Johansen]   
{Anders Johansen$^1$%
\affiliation{$^1$Leiden Observatory, Leiden University, P.O.\ Box 9513, 2300
RA Leiden, The Netherlands\break email: ajohan@strw.leidenuniv.nl}}

\pubyear{2009}
\volume{259}  
\pagerange{100--100}
\date{28/11/2008 and in revised form ??}
\jname{Cosmic Magnetic Fields: from Planets, to Stars and Galaxies}
\editors{K.G. Strassmeier, A.G. Kosovichev \& J.E. Beckman, eds.}
\begin{document}

\maketitle

\begin{abstract}

The role of magnetic fields for the formation of planets is reviewed.
Protoplanetary disc turbulence driven by the magnetorotational instability has
a huge influence on the early stages of planet formation. Small dust grains are
transported both vertically and radially in the disc by turbulent diffusion,
counteracting sedimentation to the mid-plane and transporting crystalline
material from the hot inner disc to the outer parts. The conclusion from recent
efforts to measure the turbulent diffusion coefficient of magnetorotational
turbulence is that turbulent diffusion of small particles is much stronger than
naively thought. Larger particles -- pebbles, rocks and boulders -- get trapped
in long-lived high pressure regions that arise spontaneously at large scales in
the turbulent flow. These gas high pressures, in geostrophic balance with a
sub-Keplerian/super-Keplerian zonal flow envelope, are excited by radial
fluctuations in the Maxwell stress. The coherence time of the Maxwell stress
is only a few orbits, where as the correlation time of the pressure bumps is
comparable to the turbulent mixing time-scale, many tens or orbits on scales
much greater than one scale height. The particle overdensities contract under
the combined gravity of all the particles and condense into gravitationally
bound clusters of rocks and boulders. These planetesimals have masses
comparable to the dwarf planet Ceres. I conclude with thoughts on future
priorities in the field of planet formation in turbulent discs.

\end{abstract}

\firstsection 
\section{Introduction}

Planets form in protoplanetary discs of gas and dust as the dust grains collide
and grow to ever larger bodies \cite[(Safronov 1969)]{Safronov1969}. An
important milestone is the formation of km-sized planetesimals. Drag force
interaction between particles and gas plays a big role for the dynamics of dust
particles. This way the
collisional evolution of the dust grains into planetesimals is intricately
connected to the physical state of the gas flow. The magnetorotational
instability renders Keplerian rotation profiles linearly unstable in the
presence of a magnetic field of suitable strength \cite[(Balbus \& Hawley
1991)]{BalbusHawley1991}. The ensuing magnetorotational turbulence is
currently the best candidate for driving protoplanetary disc accretion. The
relatively ease at which self-sustained magnetorotational turbulence is
produced by numerical magnetohydrodynamics codes makes it an excellent test bed
for analysing dust motion and formulating theories of planet formation in a
turbulent environment.

An interesting constraint on the magnetic field present in the solar nebula
comes from meteoritics. Most carbonaceous chondrites have a remanent
magnetisation as high as a few Gauss, frozen in as the material cooled past the
blocking temperature \cite[(Levy \& Sonett 1978)]{LevySonett1978}. A quote from
the excellent review paper by \cite{LevySonett1978} is particularly concise on
the origin of such a strong magnetic field:
\vspace{0.5cm}
\begin{quote}
``So far as we can see, there are four major candidates for the origin of the
primordial magnetic field which produced the remanence in carbonaceous
chondrites. They are:\\
\begin{enumerate}[1. ]
  \item Magnetic fields generated in very large meteorite parent bodies
  \item The interstellar magnetic field compressed to high intensity by the
  inflowing gas
  \item A strong solar magnetic field permeating the early solar system
  \item A hydromagnetic dynamo field produced in the gaseous nebula itself''
\end{enumerate}
\end{quote}
\vspace{0.5cm}
\cite{LevySonett1978} continue to put forward various physical arguments to
rule out possibility 1 and 2 [the undifferentiated parent bodies of
carbonaceous chondrites were unlikely to harbour a magnetic field, and
turbulent diffusion strongly limits the amount of field that can
be dragged into the solar nebula \cite[(Lubow et al.\ 1994)]{Lubow+etal1994}].
The magnetic field of the wind emanating from the young sun can potentially be
strong enough to imprint fields of several G at a few AU from the sun. But the
most likely scenario remains that the magnetic field was created by the
differential rotation and dynamo process in the solar nebula itself.
Simulations of magnetised shear flows indeed show that a weak seed field can be
amplified by the magnetorotational instability to a few percent of the thermal
pressure \cite[(Brandenburg et al.\ 1995, Hawley et al.\ 1996, Sano et al.\
2004)]{Brandenburg+etal1995,Hawley+etal1996,Sano+etal2004}.

In the following sections I briefly review the role of such magnetised
turbulence on the motion of dust particles and on the cosmogony of planetesimal
formation.

\section{Diffusion of small dust grains}

The magnetised turbulence in protoplanetary discs moves small dust grains
around, preventing them from sedimenting to the mid-plane and transporting
dusty material radially in the disc \cite[(Gail 2002, van Boekel et al.\ 2004)]{Gail2002,vanBoekel+etal2004}. This
section describes recent efforts to determine the turbulent diffusion
coefficient $D_{\rm t}$ of magnetorotational turbulence.

If turbulent transport can be described as a diffusion process, then the
evolution of the dust particle density $\rho_{\rm d}$ follows the partial
differential equation
\begin{equation}
  \frac{\dpa \rho_{\rm d}}{\dpa t} =
  \nab\cdot\left[D_{\rm t}\rho_{\rm g}\nab\left(\frac{\rho_{\rm
  d}}{\rho_{\rm g}}\right)\right] \, .
\end{equation}
Here $\rho_{\rm g}$ is the gas density, its presence signifying that diffusion
acts to even out differences in the solids-to-gas ratio $\epsilon_{\rm
d}=\rho_{\rm d}/\rho_{\rm g}$. The vertical flux of dust particles contains
contributions from the advection (sedimentation at velocity $w_z$) and the
diffusion,
\begin{equation}
  \mathcal{F}_z = w_z \rho_{\rm d}-D_{\rm t}\rho_{\rm g}\frac{\dpa (\rho_{\rm
  d}/\rho_{\rm g})}{\dpa z} \, .
\end{equation}
In sedimentation-diffusion equilibrium we have $\mathcal{F}_z=0$. Setting the
velocity of the dust particles its terminal value, $w_z=-\tau_{\rm
f}\varOmega^2 z$ (where $\varOmega$ is the Keplerian frequency and $\tau_{\rm
f}$ is the friction time of the particles), gives the solution \cite[(e.g.\
Dubrulle et al.\ 1995)]{Dubrulle+etal1995}
\begin{equation}
  \epsilon_{\rm d}(z)=\epsilon_1 \exp[-z^2/(2 H_\epsilon^2)]
\end{equation}
for the solids-to-gas ratio $\epsilon_{\rm d}=\rho_{\rm d}/\rho_{\rm g}$. The
scale height $H_\epsilon$ follows the expression
\begin{equation}
  H_\epsilon^2=\frac{D_{\rm t}}{\tau_{\rm f}\varOmega^2} \, ,
  \label{eq:heps}
\end{equation}
while the solids-to-gas ratio in the mid-plane is given by
\begin{equation}
  \epsilon_1=\epsilon_0\sqrt{\left(\frac{H}{H_\epsilon}\right)^2+1} \, .
\end{equation}
Here $H$ is the pressure scale height of the gas.
In the above derivations we have assumed (a) that the friction time is
independent of height over the mid-plane and (b) that the diffusion coefficient
is independent of height over the mid-plane. None of these assumptions are true
in general, but if we stay within a few scale heights of the mid-plane and
treat the diffusion coefficient as a suitably averaged diffusion coefficient,
then the expressions are relatively good approximations.

\begin{figure}
  \begin{center}
    \includegraphics[width=0.5\linewidth]{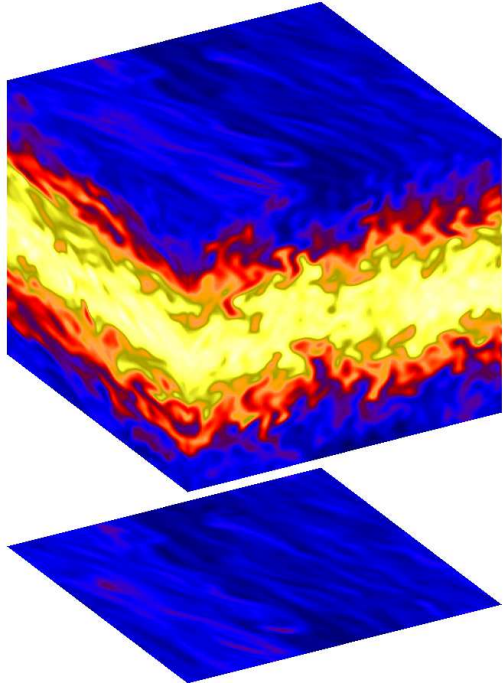}
  \end{center}
  \caption{The dust density at the sides of a simulation box corotating with
  the disc at an arbitrary distance from the central star. The
  radial direction points right, the azimuthal direction left and up, while the
  vertical direction points directly up. The dust density distribution arises
  from an equilibrium between sedimentation and turbulent diffusion by the
  magnetorotational turbulence.}
  \label{f:dust_box}
\end{figure}
In a real turbulent flow the observed diffusion-sedimentation equilibrium can
be used to measure the turbulent diffusion coefficient of the flow. In
figure~\ref{f:dust_box} we show an example of such a diffusion-sedimentation
equilibrium \cite[(from Johansen \& Klahr 2005)]{JohansenKlahr2005}
for a shearing box simulation of magnetorotational turbulence. The problem
of determining the diffusion coefficient is thus
reduced to measuring the scale height $H_\epsilon$ of the dust in
figure~\ref{f:dust_box}. Using equation~\ref{eq:heps} then directly
yields a value of $D_{\rm t}$. Obviously the diffusion coefficient must scale
with the overall strength of the turbulence. The interesting quantity to
determine is thus the Schmidt number ${\rm Sc}$, defined as the turbulent
viscosity coefficient relative to the turbulent diffusion coefficient, ${\rm
Sc}=\nu_{\rm t}/D_{\rm t}$. In a Keplerian disc the turbulent viscosity is in
turn defined from the Reynolds and Maxwell stresses,
\begin{equation}
  \nu_{\rm t} = \frac{2}{3}\frac{\langle \rho u_x u_y -
  \mu_0^{-1} B_x B_y \rangle}{\langle\rho\rangle}
\end{equation}
The Schmidt number was found by \cite{JohansenKlahr2005} to be around $1.5$ for
vertical diffusion and $0.85$ for radial diffusion. This is surprisingly close
to unity and a bit mysterious given that the turbulent viscosity is dominated
by the magnetic Maxwell stress $\langle -\mu_0^{-1} B_x B_y \rangle$. This
stress does not
directly affect the dust particles. A possible explanation is that diffusion is
determined by the diagonal entries in the $u_i u_j$ correlation tensor. These are
much higher than the off diagonal Reynolds stress $u_x u_y$. Thus the MRI
inherently transports a passive scalar (by fluid motion) and the angular
momentum (by magnetic tension) equally well.

Different groups have used various independent methods to measure the turbulent
diffusion coefficient of magnetorotational turbulence. A vertical Schmidt
number of around unity was measured by \cite{Turner+etal2006}, while
\cite{FromangPapaloizou2006} reported a value of approximately three. This
gives some confidence that the Schmidt number is well constrained.
However, \cite{Carballido+etal2005} found a radial Schmidt number as high as
ten in relatively strong turbulence. To address the discrepancy between this
value and the much lower value found by \cite{JohansenKlahr2005},
\cite{JohansenKlahrMee2006} performed simulations of the MRI with various
strengths of an imposed, external field, yielding a higher turbulent viscosity
than in zero net flux simulations. The Schmidt number was indeed found to
decrease with increasing strength of the turbulence. Stronger turbulence (such
as in Carballido et al.\ 2005) is less good at diffusing dust particles
relative to its stresses. The explanation is that the correlation time of
the turbulence decreases with increasing turbulent energy, and that turbulent
structures do not stay coherent long enough to effectively diffuse particles.

Large particles partially decouple from the turbulence and are primarily
diffused by large scale eddies with relatively long correlation times. The
experiments by \cite{Carballido+etal2006} indeed showed that the diffusion
coefficient falls rapidly for particles above a few metres in size, in good
agreement with the analytical derivations of \cite{YoudinLithwick2007}.

\section{Zonal flows}

While smaller dust particles are clearly prevented from forming a very thin
mid-plane layer by the magnetorotational turbulence, pebbles, rocks and
boulders begin to gradually decouple from the gas. Accretion discs are radially
stratified with a pressure that decreases with distance from the star. The
pressure gradient acts to reduce the effect of gravity felt by the gas, and as
a result the gas rotates slightly slower than Keplerian. The particles,
however, do not react to gas pressure gradients and aim to orbit with the
Keplerian speed. The head wind of the slower rotating gas drains the particles
of angular momentum and they spiral towards the star in a few hundred orbital
periods \cite[(Weidenschilling 1977)]{Weidenschilling1977}.

\begin{figure}
  \begin{center}
    \includegraphics[width=\linewidth]{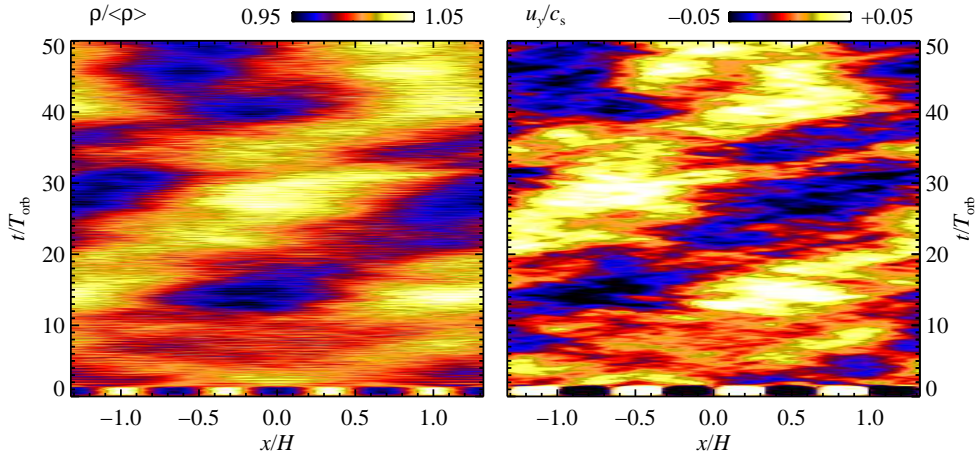}
  \end{center}
  \caption{The gas density (left plot) and the azimuthal velocity (right plot)
  as a function of the radial distance from the centre of the box, $H$, and the
  time, $t$, measured in orbits. There is a perfect $-\pi/2$ phase difference
  between the pressure bump and the zonal flow, in agreement with a geostrophic
  balance. The zonal flow has in turned been excited by a large scale variation
  in the Maxwell stress.}
  \label{f:uymx_t_nu2}
\end{figure}

The radial pressure profile of gas in turbulent discs need not be monotoneously
falling. The presence of large scale, long-lived pressure bumps leads to
concentrations of migrating dust particles into radial bands. Simulations of
magnetorotational turbulence in a box gives evidence that such pressure bumps
form spontaneously in the turbulent flow \cite[(Johansen et al.\ 2006, Johansen
et al.\ 2009)]{JohansenKlahrHenning2006,Johansen+etal2009}. In
figure~\ref{f:uymx_t_nu2} we plot the gas density and the azimuthal velocity,
averaged over the azimuthal and vertical directions, as a function of radial
distance from the centre of the box $x$ and the time $t$. The gas density
exhibits axisymmetric column density bumps with amplitude around 5\% of the
average density. These bumps are surrounded by a sub-Keplerian/super-Keplerian
zonal flow, maintaining perfect geostrophic balance with $2 \rho_0 \varOmega
u_y\approx\dpa P/\dpa r$.

Varying resolution, presence or non-presence of stratification, dissipation
parameters and dissipation types, \cite{Johansen+etal2009} find that
pressure bumps and zonal flows are ubiqituous in shearing box
simulations of magnetorotational turbulence,
provided that the simulation box is large enough (more than one scale height in
radial extent) and possibly also that the physical dissipation is high enough.
What is then the launching mechanism for these zonal flows? Large
scale fluctuations in the Maxwell stress lead to a differential transport of
momentum. Thus the magnetic field is responsible for separating the orbital
flow into regions of slightly faster and slightly slower rotation.

A model of the excitation of zonal flows and pressure bumps can be obtained
from a simplified version of the dynamical equations,
\begin{eqnarray}
  0 &=& 2 \varOmega \hat{u}_y - \frac{c_{\rm s}^2}{\rho_0} \ii k_0 \hat{\rho}
  \, , \\
  \frac{\de \hat{u}_y}{\de t}  &=& -\frac{1}{2} \varOmega \hat{u}_0 + \hat{T}
  \, , \\
  \frac{\de \hat{\rho}}{\de t} &=& -\ii k_0 \hat{u}_x -
  \frac{\hat{\rho}}{\tau_{\rm mix}} \, .
\end{eqnarray}
Here $\hat{u}_x$, $\hat{u}_u$ and $\hat{\rho}$ are the amplitudes of the radial
and azimuthal velocity and gas density at the largest radial scale of the
simulation, with wavenumber $k_0$. The first equation denotes geostrophic
balance, while we have kept the time evolution terms in the two other
equations. Non-linear terms enter through $\hat{T}$, the large scale magnetic
tension, and $\hat{\rho}/\tau_{\rm mix}$, turbulent diffusion of the mass
density.
\begin{figure}
  \begin{center}
    \includegraphics{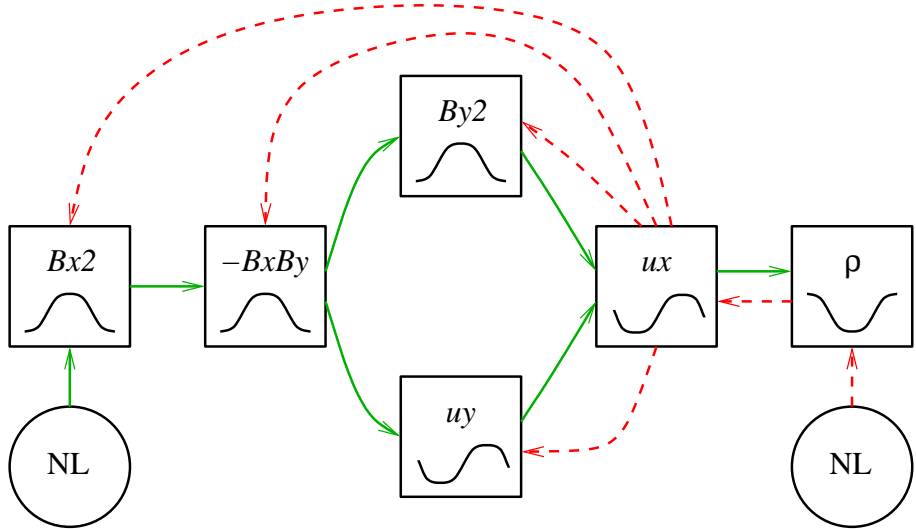}
  \end{center}
  \caption{Diagram of how non-linear excitation of the large scale radial
  magnetic energy leads to the excitation of zonal flow. Green arrows label
  positive energy transfer, while red arrows (dashed) denote energy sinks.
  Non-linear interactions are responsible both for the excitation and for the
  balance, the latter through diffusive mixing of the gas density.}
  \label{f:confusogram}
\end{figure}

We can combine the above equations into a single evolution equation for the
density,
\begin{equation}\label{eq:drhohatdt}
  \frac{\de \hat{\rho}}{\de t} = \frac{1}{1+ k_0^2 H^2 } \left( \hat{F}
      - \frac{\hat{\rho}(t)}{\tau_{\rm mix}} \right) \, ,
\end{equation}
where $ \hat{F} \equiv - 2 \ii k_0 \rho_0\hat{T}/\varOmega$ is the forcing
term. The prefactor $c_k\equiv(1 + k_0^2 H^2)^{-1}$ is a pressure correction
for small-scale modes that both decreases the amplitude of the forcing and
increases the effective damping time. The coherence time-scale of the Maxwell
stress (and thus of $\hat{F}$), $\tau_{\rm for}$, is generally much shorter
than the mixing time-scale, $\tau_{\rm mix}$. Thus we need to model
equation~\ref{eq:drhohatdt} as a stochastic differential equation \cite[(see
e.g.\ Youdin \& Lithwick 2008)]{YoudinLithwick2008}. The
Maxwell stress gives short, uncorrelated kicks to the zonal flow. This would
lead to an amplitude that grows as the square root of time. However, in
presence of turbulent diffusion, the solution tends to
\begin{equation}
  \frac{\hat{\rho}_{\rm eq}}{\rho_0} = 2 \sqrt{c_k \tau_{\rm for}\tau_{\rm mix}}
    H k_0 \frac{\hat{T}}{c_{\rm s}} \, .
\end{equation}
The correlation time of the zonal flows is predicted to be equal to the mixing
time-scale, in good agreement with the results. The model also predicts that
$\hat{\rho}_{\rm eq} \propto k^{-2}$ for $k_0 H\gg 1$. This is in very good
agreement with the very clearly sinusoidal density fluctuations seen in
figure~\ref{f:uymx_t_nu2}.

The cause of the large scale variation in the Maxwell stress remains unknown.
\cite{Johansen+etal2009} argue that magnetic energy takes part in an inverse
cascade from the moderate scales, excited directly by the MRI, to large scales.
A diagram of the zonal flow excitation appears in figure~\ref{f:confusogram}.
Note that the above zonal flow excitation model assumes that the magnetic
tension (i.e. $-B_xB_y$ in figure~\ref{f:confusogram}) is
given, whereas in fact one may go on step further back to $B_x^2$, which is
excited directly by a non-linear term. The Maxwell stress then increases from
the Keplerian stretching of the radial field. The model also predicts that the
magnetic pressure should grow in anti-phase with the thermal pressure. This is
indeed also observed.

\section{Planetesimal formation}

\begin{figure}
  \begin{center}
    \includegraphics{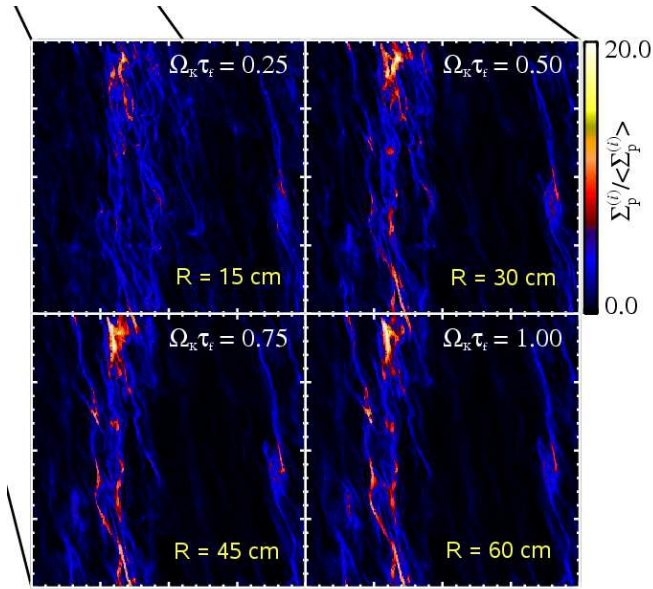}
  \end{center}
  \caption{The column density of four different particle sizes, before
  self-gravity has been switched on. The particles concentrate at the same
  locations, but larger particles experience a higher local column density.}
  \label{f:before_selfgravity}
\end{figure}
The zonal flows presented in the last section are very efficient at trapping
particles. At the outer sub-Keplerian side the particles face a slightly
stronger headwind and drift faster inwards. At the inner super-Keplerian side
the particles experience a slight backwind and move out. The effect of pressure
bumps on the migration of rocks and boulders goes at least back to
\cite{Whipple1972}. It has later received extensive analytical treatment by
\cite{KlahrLin2001} and by \cite{HaghighipourBoss2003}. The narrow box
simulations of \cite{HodgsonBrandenburg1998} found no evidence for long-lived
concentrations of relatively tighly coupled particles in magnetorotational
turbulence. However, \cite{JohansenKlahrHenning2006} observed concentrations of
marginally coupled dust particles (cm-m sizes), by up to two orders of
magnitude higher than the average paricle density, in high pressure regions
occuring in magnetorotational turbulence. In a simulation of a (part of a)
global disc
\cite{FromangNelson2005} reported similar concentrations in a long-lived vortex
structure.

The question of how long-lived high pressure structures form and survive in
magnetised turbulence is of general interest. However, their effect on
planetesimal formation is no less intriguing. \cite{Johansen+etal2007} expanded
earlier models of boulders in turbulence by considering several particle sizes
simultaneously and solving for the self-gravity of the boulders. First the
turbulence is allowed to develop for 20 local rotation periods without the
gravity of the particles (which is weak anyway). This way a sedimentary
mid-plane layer, with a width of a few percent of the gas scale height, forms
in equilibrium between sedimentation and turbulent diffusion. In
figure~\ref{f:before_selfgravity} we show the column density of the four
different particle sizes -- rocks and boulders with sizes 15 cm, 30 cm, 45 cm,
and 60 cm. A weak zonal flow has been sufficient to create bands of very high
particle overdensity. An additional instability in the coupled motion of gas
and dust has further augmented the local overdensities \cite[(Goodman \&
Pindor 2000, Youdin \& Goodman 2005, Youdin \& Johansen 2007, Johansen \&
Youdin
2007)]{GoodmanPindor2000,YoudinGoodman2005,YoudinJohansen2007,JohansenYoudin2007}.

\begin{figure}
  \begin{center}
    \includegraphics[width=0.7\linewidth]{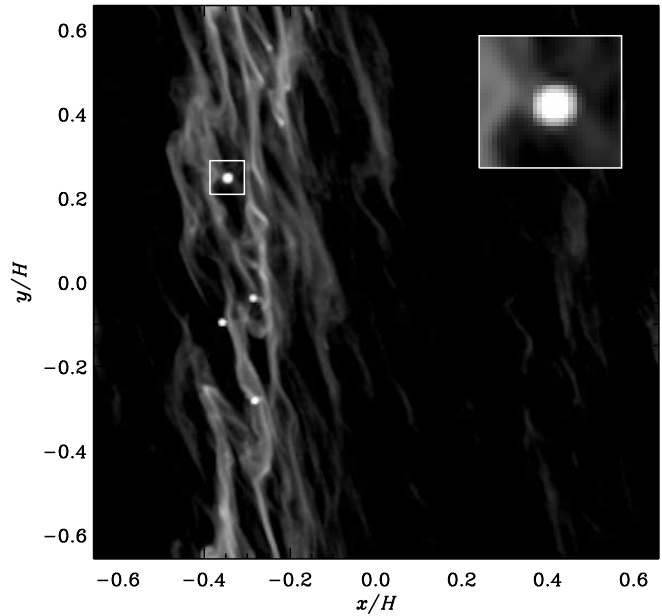}
  \end{center}
  \caption{The column density at $\Delta t=7 T_{\rm orb}$ after self-gravity is
  turned on. Four gravitationally bound clusters of rocks and boulders have
  condensed out of the flow. The most massive cluster (see enlargement) has a
  mass comparable to the dwarf planet Ceres by the end of the simulation.}
  \label{f:planetesimals_bw}
\end{figure}

As the self-gravity of the disc is activated, the overdense bands contract
radially. Upon reaching the local Roche density, a full non-axisymmetric
collapse occurs and a few gravitationally bound clusters of rocks and boulders
condense out of the particle layer. The
column density of the particles is shown in figure~\ref{f:planetesimals_bw}.

\section{Conclusions}

The presence of magnetic fields in protoplanetary discs is of vital importance
for planet formation and for observational properties of protoplanetary discs.
Small dust grains are transported very efficiently by the turbulence. While
this counteracts sedimentation to the mid-plane, and thus prevents the
razor thin mid-plane layer of \cite{GoldreichWard1973} from forming, the
turbulent transport underlies the presence of small dust grains many scale
heights above the disc mid-plane. The presence of crystalline silicates in the
cold outer regions of discs \cite[(Gail 2002, van Boekel et al.\
2004)]{Gail2002,vanBoekel+etal2004} can likely also be attributed to turbulent
diffusion \cite[(but see Dullemond et al.\ 2006 for an alternative
view taking into account disc formation history)]{Dullemond+etal2006}.

Larger dust particles -- pebbles, rocks, and boulders -- slow down or reverse
the radial migration as they encounter variations in the radial pressure
gradient. Fluctuations in the Maxwell stress, with a coherence time of a few
orbits, launch axisymmetric zonal flows. These flows in turn go into
geostrophic balance with a radial pressure bump. The concentrations of solid
particles in such pressure ridges can get high enough for a gravitational
collapse into planetesimals to occur. However, a satisfactory mechanism for
setting the scale of the pressure bumps is lacking, as the bumps grow to fill
the box for all considered box sizes in \cite{Johansen+etal2009}. The final
size may ultimately be set by global curvature effects \cite[(Lyra et al.\
2008a)]{Lyra+etal2008a}.

An important problem related to the motion of dust particles in turbulence is
their collision speeds. The relative speed of small particles approaches zero
as the particle separation is decreased. But particles that are only marginally
coupled to the turbulent eddies have a significant memory of their trajectories
and can collide at non-zero speeds. \cite{Carballido+etal2008} indeed found
that the relative speeds of large particles is unchanged below a certain
separation, giving confidence that the proper collision speed has been found.
Turbulent eddies with sizes around the stopping length of the particle are most
efficient at inducing relative motion. However, even for marginally coupled
particles these eddies may be on the edge of the dissipative subranges of the
turbulence, due to the limited resolution of numerical simulations.
\cite{Johansen+etal2007} found that the collision speed, measured as the
velocity difference over a single grid cell, increases by 10--20\% each time
the resolution is doubled. Ultrahigh resolution measurements of large scale and
short scale relative speeds, and comparison to analytical models \cite[(V\"olk
et al.\ 1980, Cuzzi et al.\ 1993, Schr\"apler \& Henning 2004, Youdin \&
Lithwick
2008)]{Voelk+etal1980,Cuzzi+etal1993,SchraeplerHenning2004,YoudinLithwick2007}
and to sticking experiments \cite[(Wurm et al.\ 2006, Blum \& Wurm
2008)]{Wurm+etal2006,BlumWurm2008}, is an important future priority for our
picture of how planets form in turbulent gas discs.

To our best knowledge parts of the solar nebula had so low ionisation fraction
that the collisional resistivity was too high for the magnetorotational
instability to develop \cite[(e.g. Gammie 1996, Sano et al.\
2000)]{Gammie1996,Sano+etal2000}. \cite{KretkeLin2007} and
\cite{Brauer+etal2008} modelled the sharp increase in resistivity as the gas
temperature drops below the freezing point of ice (the so-called snow line) at
a few AU from the sun.
The corresponding drop in turbulence activity leads to a pile up of gas and a
run away growth in particle density from the influx of migrating solid
particles. The lack of radial drift in such a location allows for planetesimal
formation by coagulation to occur without having to compete with the radial
drift time-scale. The edges of ``dead zones'' can be unstable to a Rossby wave
instability \cite[(Inaba \& Barge 2006, Varni\`ere \& Tagger
2006)]{InabaBarge2006,VarniereTagger2006}. The Rossby vortices efficiently trap
particles, which leads to a burst of planet formation at the edge of dead zones
\cite[(Lyra et al.\ 2008b)]{Lyra+etal2008b}. This way magnetic fields helps the
planet formation process both in their presence and in their absence.

\begin{acknowledgments}
  I would like to thank my collaborators Andrej Bicanski, Andrew Youdin, Axel
  Brandenburg, Frithjof Brauer, Hubert Klahr, Jeff Oishi, Kees Dullemond,
  Mordecai-Mark Mac Low, Thomas Henning, and Wladimir Lyra.
\end{acknowledgments}

\begin{discussion}

\end{discussion}

\end{document}